\documentclass[twocolumn,aps,prl,showpacs]{revtex4}
\usepackage{epsfig}
\usepackage{color}

\begin{document}


\title{Spatial modulation of a unitary impurity-induced resonances in superconducting CeCoIn$_{5}$}
\author{Ge Zhang, Bin Liu}

\affiliation{Department of Physics, Beijing Jiaotong University, Beijing 100044, China}
\author{Yi-feng Yang }
\affiliation{Beijing National Laboratory for Condensed Matter Physics and\\
Institute of Physics, Chinese Academy of Sciences, Beijing 100190, China}
\affiliation{Collaborative Innovation Center of Quantum Matter, Beijing 100190, China}

\author{Shiping Feng}
\affiliation{Department of Physics, Beijing Normal University, Beijing 100875, China}
\begin{abstract}

Motivated by recent experimental progress in high-resolution scanning tunneling microscopy (STM) techniques, we propose to investigate the local quasiparticle density of states around a unitary impurity in the heavy fermion superconductor CeCoIn$_{5}$. Based on the T-matrix approach we obtain a sharp nearly zero-energy resonance state (ZERS) in the strong impurity potential scattering localized around the impurity, and find qualitative differences in the spatial pattern of the tunneling conductance modulated by the nodal structure of the superconducting gap. These unique features may be used as a probe of the superconducting gap symmetry and in combination with the further STM measurements, may help to confirm the $d_{x^{2}-y^{2}}$ pairing in CeCoIn$_{5}$ at ambient pressure.

\end{abstract}
\pacs{74.25.Jb, 74.20.Pq, 74.50.+r, 74.62.En}

\maketitle

Superconducting gap symmetry reflects the nature of the underlying attractive interactions that give rise to the Cooper pairs in unconventional superconductors \cite{Scalapino}. Tremendous efforts have been devoted to the measurement of their gap structures in the past decades. Experimental detection of the paring symmetry has proven difficult. Among the many techniques that have been developed, the angle-resolved photoemission spectroscopy (ARPES) provides an exact mapping of the band structures but in many cases it is limited by the energy resolution \cite{Fischer}. This is in particular the case for heavy fermion superconductors, which typically has a superconducting transition temperature of a few Kelvin and a superconducting gap of $\sim1\,$meV and is so far indiscernible in ARPES measurment \cite{White,Knebel,Stewart}. Other techniques such as the specific heat measurement, the nuclear magnetic/quadrupole resonance (NMR/NQR) and the neutron scattering measurement can only provide indirect evidences for the gap structure. As a result, the pairing symmetry in many heavy fermion superconductors remain undetermined \cite{White,Knebel}. One noticeable example is CeCu$_{2}$Si$_{2}$, which is the first unconventional superconductor discovered in history and has long been thought to be a $d$-wave superconductor at ambient pressure \cite{Steglich}. However, recent specific heat  measurement down to very low temperature suggests that it may instead have two nodeless gaps \cite{Kittaka}; theoretical calculations afterwards suggest that superconductivity might actually be $s^\pm$-wave \cite{Ikeda1}.

In this work, we focus on CeCoIn$_5$ which belongs to the famous Ce-115 family, CeMIn$_{5}$ (M = Co, Rh, Ir) \cite{Hegger,Petrovic0,Petrovic}, and has the highest T$_{c}$$\approx2.3$\,K in all Ce-based superconductors at ambient pressure. This family share many similarities with the cuprate superconductors as well as the newly discovered pnictide superconductors, for example, the quasi-2D Fermi surfaces (FS) \cite{McCollam} and the competition between antiferromagnetism (AF) and superconductivity \cite{Shang,Morr,Allan,Zhou,Hu,Kenzelmann,Lu,Park}. It is hence generally believed that superconductivity in CeCoIn$_5$ is driven by strong AF spin-fluctuations \cite{Morr,Allan,Zhou,Hu,Kenzelmann}. Many experiments and theoretical calculations indeed have pointed to an unconventional d-wave pairing \cite{Shang,Morr,Allan,Zhou,Hu,Kenzelmann,Lu,Park,Kohori,Matsuda,Izawa,Aoki,An,Ikeda,Stock,Eremin,Hiasa,Park1,Vorontsov,Ronning,Das,Bin}. However, the exact gap structure, either $d_{x^{2}-y^{2}}$ or $d_{xy}$, has so far not been underpinned, despite that it has been generally believed to be $d_{x^2-y^2}$. Early thermal conductivity and specific heat experiments have produced very controversial results concerning its $d_{x^{2}-y^{2}}$ or $d_{xy}$ symmetry \cite{Matsuda,Izawa,Aoki}. Recent field-angle-resolved specific heat measurement \cite{An} observed a puzzling change of symmetry from $d_{xy}$ to $d_{x^{2}-y^{2}}$ with decreasing temperature, which was, not without uncertainty, attributed to the so-called Doppler shift of the nodal quasiparticles \cite{Vorontsov}. On the other hand, anisotropy in the high-field superconducting phase seems to favor the $d_{xy}$ symmetry \cite{Ikeda}, but the existence of a magnetic resonance in the neutron scattering experiment supports the $d_{x^{2}-y^{2}}$ symmetry \cite{Stock,Eremin}. To the best of our knowledge, APRES has not been able to detect the superconducting gap \cite{Koitzsch} and phase sensitive measurements are still lacking.

Very recently, new spectroscopic technique was developed based on high-resolution scanning tunneling microscopy (STM/STS) \cite{Aynajian}. It shows that quasiparticle interference (QPI) due to impurity scattering of $f$-electrons may lead to spatial patterns in the conductance spectrum that could be used to determine the quasiparticle band dispersions \cite{Morr,Allan,Zhou,Ernst}. In this work, we extend this idea and use the T-matrix approach to investigate the spatial modulation of the local electronic structures around a unitary nonmagnetic impurity in the superconducting phase. This method has been quite successful in identifying the pairing states in other unconventional superconductors \cite{Zhu}. We extend it to CeCoIn$_{5}$ based on the recent STM experiment \cite{Morr,Allan,Zhou} and propose an alternative way to distinguish the $d_{x^{2}-y^{2}}$ or $d_{xy}$ symmetry of the superconducting gap structure.

The quasiparticle band structures have been analyzed in the STM experiment and found in good agreement with the first-principles band structure calculations \cite{Maehira,Onuki,Haga,Hall,Shishido}. The effective low-energy Hamiltonian for the emergent heavy quasiparticles can be written as \cite{Tanaka},
\begin{eqnarray}
H&=&\sum_{\bf k,\sigma}\varepsilon^{c}_{\bf k}c^{\dag}_{\bf k,\sigma}c_{\bf k,\sigma}+\sum_{\bf k,\sigma}E^{f}_{\bf k}f^{\dag}_{\bf k,\sigma}f_{\bf k,\sigma}\nonumber\\&+&\sum_{\bf k,\sigma}V_{\bf k}f^{\dag}_{\bf k,\sigma}c_{\bf k,\sigma}+h.c.
\end{eqnarray}
where $E^{f}_{\bf k}$ and $\varepsilon^{c}_{\bf k}$ are the effective dispersions of the $f$-band and the conduction band, respectively, and $V_{\bf \bf k}$ is the effective hybridization strength renormalized by the strong on-site Coulomb repulsion of the $f$-electrons. The above effective low-energy Hamiltonian can be diagonalized to give two hybridized quasiparticle bands,
\begin{eqnarray}
 E^{\alpha,\beta}_{\bf k}&=&\frac{1}{2}\left[(\varepsilon^{c}_{\bf k}+E^{f}_{\bf k})\pm\sqrt{(E^{f}_{\bf k}-\varepsilon^{c}_{\bf k})^{2}+4V^{2}_{\bf k}}\right].
\end{eqnarray}
The compound of CeCoIn$_{5}$ has a very complicated 3D FS. Tight binding models often approximate the FS by a large cylindrical electron pocket centered at $(\pi,\pi,k_{z})$ and a hole like one centered at $\Gamma$. In this case, the 2D band structure has been approximately used to investigate the unconventional superconductivity of CeCoIn$_{5}$\cite{Knebel,Morr,Allan,Zhou,Eremin}. As shown in Fig. 1(a), fitting to the experimental QPI dispersion \cite{Allan,Zhou} yields the FS topology (the details form of the band structure and effective hybridization strength are given in Ref. 15), which consists of two Fermi sheets (denoted by $\alpha$ and $\beta$) in good agreement with de Haas-van Alphen (dHvA) experiments \cite{McCollam,Haga,Hall,Shishido} and ARPES measurements \cite{Dudy}. The calculated band structures in Fig. 1(b) display a sharp van Hove singularity at the X point of the $\alpha$-band, so the system can be very susceptible to correlation effects. In addition, we note that the $\alpha$-band (red line in Fig. 1(c)) has a much larger density of state near the Fermi energy and should dominate the low energy properties, although both bands contribute to the Fermi surfaces in CeCoIn$_{5}$.

\begin{figure}[tbp]
\begin{center}
\includegraphics[width=1.0\linewidth]{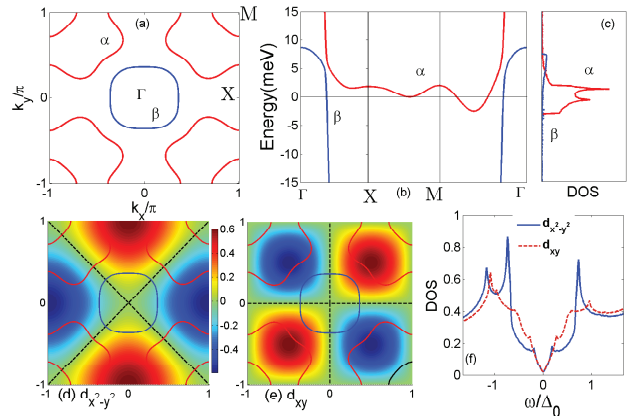}
\end{center}
\caption{(Color online) (a) Calculated Fermi surface for CeCoIn$_{5}$. (b) The corresponding band structure. (c) The normal state density of state. Visualization of the overlap between FS (solid line) and gap functions for (d) $d_{x^{2}-y^{2}}$ and (e) $d_{xy}$ pairing symmetries. The dash-line denotes the nodal line. (f) The density of states in superconducting state for two gap symmetries.}
\end{figure}

To study the superconducting properties, we introduce a four-component Nambu spinor operator, $\psi^{\dag}_{\bf k}=(a^{\dag}_{\beta,{\bf k}\uparrow},a_{\beta,-{\bf k}\downarrow},a^{\dag}_{\alpha,{\bf k}\uparrow},a_{\alpha,-{\bf k}\downarrow})$. The matrix Green's function in the superconducting state is then given by
\begin{eqnarray}
\hat{G}^{-1}_{0}({\bf k},i\omega_{n})=i\omega_{n}\hat{1}-\left (\matrix{E^{\beta}_{\bf k} &\Delta^{\beta}_{\bf k}
&0 &0\cr \Delta^{\beta}_{\bf k} &-E^{\beta}_{\bf k}
&0 &0\cr 0 &0 &E^{\alpha}_{\bf k} &\Delta^{\alpha}_{\bf k}\cr 0 &0 &\Delta^{\alpha}_{\bf k} &-E^{\alpha}_{\bf k}\cr}\right),
\end{eqnarray}
in which two different gap structures, $d_{x^{2}-y^{2}}$ with $\Delta^{\alpha,\beta}_{\bf k}=\Delta_{0}(\cos k_{x}-\cos k_{y})/2$ and $d_{xy}$ with $\Delta^{\alpha,\beta}_{\bf k}=\Delta_{0}(\sin k_{x}\sin k_{y})$, will be discussed in the following for CeCoIn$_5$. $\Delta_{0}=0.6meV$ is the magnitude of the d-wave gap and is determined from experiment \cite{Allan}. Figs. 1(d) and (e) compare the FS topology and the gap structures in the first Brillouin zone. For $d_{x^{2}-y^{2}}$ (Fig. 1(d)), the nodal lines cross all Fermi surfaces enclosed by $\alpha$ and $\beta$-bands, giving rise to a sign change within each pocket so the calculated DOS, as shown in Fig. 1(e) and probed by STM \cite{Morr,Allan,Zhou,Ernst}, exhibits a typical V-shape near the Fermi energy. Whereas for $d_{xy}$ (Fig. 1(e)), one sees an immediate distinction: the nodal lines in the superconducting gap do no intersect with the dominant $\alpha$-band so that there is no sign change within each electron pocket enclosed by the $\alpha$-band. However, the gap amplitude is very small on the Fermi sheets near the M-point. As a result, the local DOS also exhibits a V-shape around the Fermi energy which makes it hard to be distinguished from that of the $d_{x^2-y^2}$-wave gap in the STM experiment. Although the superconducting coherence peaks near the gap edges are also noticeably more sharp for $d_{x^2-y^2}$, these quantitative differences could easily be smeared out by correlation effects or thermal broadening and cannot be relied on as an experimental indicator. It is necessary to have some properties that can be qualitatively distinguished for the STM measurements.

Motivated by the recent QPI experiment, here we examine the perturbation to the local electronic structures in response to local impurities for different superconducting gap symmetry. The effect of the impurity scatterings can be treated within the $T$-matrix approach. For simplicity, we consider only a single nonmagnetic impurity and use the  scattering matrix, $\hat{U}=\tau_{0}\otimes u\sigma_{0}+\tau_{x}\otimes v\sigma_{0}$, in the Nambu representation. Here $u$ and $v$ are the strength of the intra- and inter-band scattering potential, and $\tau_{i}$ and $\sigma_{i}$ are the Pauli matrices acting in spin and orbital space, respectively. This yields the full interacting Green's function,
\begin{eqnarray}
\hat{G}({\bf i},{\bf j},i\omega_{n})&=&\hat{G}_{0}({\bf i-j},i\omega_{n})\nonumber\\&+&\hat{G}_{0}({\bf i},i\omega_{n})\hat{T}\hat{G}_{0}(-{\bf j},i\omega_{n}),
\end{eqnarray}
where $\hat{G}_{0}({\bf j},i\omega_{n})=\frac{1}{N}\sum_{\bf k}e^{i\bf k\cdot{\bf j}}\hat{G}_{0}({\bf k},i\omega_{n})$ is the the real-space bare Green's function
and $\hat{T}(i\omega_{n})=\hat{U}[\hat{1}-\hat{G}_{0}({\bf 0},i\omega_{n})\hat{U}]^{-1}$ is the the $T$-matrix that has incorporated all the scattering processes.

\begin{figure}[tbp]
\begin{center}
\includegraphics[width=1.05\linewidth]{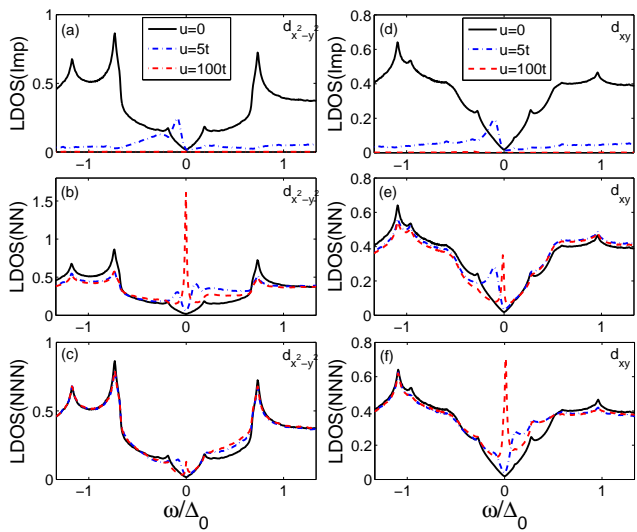}
\end{center}
\caption{(Color online) The LDOS spectra of quasiparticles as a function of energy by varying intra-band scattering potential strength $u$. The left panels indicate the pairing state with $d_{x^{2}-y^{2}}$ for a) impurity site (Imp), b) the nearest neighboring site (NN), and c) the next-nearest neighboring site (NNN), respectively. The right panels are the same with the left ones except for the pairing state with $d_{xy}$.}
\end{figure}

To compare with STM experiment, we carry out the analytic continuation $i\omega_{n}\rightarrow\omega+i0^{\dagger}$ for the full Green's function and calculate the local density of states (LDOS) in real space:
\begin{eqnarray}
\rho({\bf i},\omega)=-\frac{1}{\pi}{\rm Im} {\rm Tr} [\hat{G}({\bf i},{\bf i},\omega+i0^{\dagger})].
\end{eqnarray}
For simplicity, we first assume $v=0$ and consider only the intra-band impurity scattering $u$. The results are qualitatively the same even in the presence of inter-band impurity scatterings (not shown here).
Fig. 2 plots the calculated DOS for $d_{x^{2}-y^{2}}$ and $d_{xy}$ at the impurity site (Imp), the nearest-neighbor (NN) sites and the next-nearest-neighbor (NNN) sites. For comparison, the black solid lines show the LDOS in the clean system ($u=v=0$). One immediately sees that the DOS at the impurity site is strongly suppressed compared to the clean system. This could be easily understood because the impurity potential pushes the local f-electron level to a very high energy and effectively depletes the local site. For all other sites, the LDOS is not much affected, except that one observes emergence of sharp resonance peaks inside the superconducting gap. These resonances originate from the poles in the full Green's function caused by the impurity scattering. Taking the $\alpha$ band as an example, the impurity correction to the Green's function, $\delta \hat{G}=\hat{G}-\hat{G}_0$, is given by
\begin{eqnarray}
\delta \hat{G}^{33}({\bf r},{\bf r},i\omega_{n})=\frac{u[\hat{G}^{33}_{0}({\bf r},i\omega_{n})]^{2}}{1-u\hat{G}^{33}_{0}({\bf 0},i\omega_{n})}-\frac{u[\hat{G}^{34}_{0}({\bf r},i\omega_{n})]^{2}}{1-u\hat{G}^{33}_{0}({\bf 0},-i\omega_{n})},
\end{eqnarray}
where ${\bf r}$ is the coordinates relative to the impurity site, $\hat{G}^{33}_{0}$ and $\hat{G}^{34}_{0}$ are the normal and anomalous Green's functions in the SC state, respectively. The poles are then determined by $u^{-1}=\textrm{Re}\hat{G}^{33}_{0}({\bf 0},\pm\omega_{res})$, which gives rise to sharp resonances at either negative (particle-like, $\omega_{res}<0$) or positive (hole-like, $\omega_{res}>0$) energy. The spectral weights of these resonance peaks are determined by the prefactors, $[\hat{G}^{33}_{0}({\bf r},i\omega_{n})]^{2}$ or $[\hat{G}^{34}_{0}({\bf r},i\omega_{n})]^{2}$.

We first discuss $d_{x^2-y^2}$-wave. The results are plotted in Figs. 2(a-c). While the LDOS is strongly suppressed at the impurity site, we still see a very weak resonance peak at the negative energy. But the resonance at positive energy is missing. Based on the above formula, we can understand this from the anomalous part of the Green's function $\hat{G}^{34}_{0}({\bf r},i\omega_{n})$ which is roughly the Fourier transformation of the gap function. Hence for any $d$-wave gap, the sum over k-space cancels out at the local site ${\bf r}=(0,0)$ so the positive energy peak is missing. For the nearest-neighbor site [{\bf r}=(1,0)], as is shown in Fig. 2(b), the k-space summation indeed gives rise to two resonance peaks at both positive and negative energies. Moreover, as show in Figs. 3(a) and (c), we have $[\hat{G}^{34}_{0}({\bf r},i\omega_{n})]^{2}>[\hat{G}^{33}_{0}({\bf r},i\omega_{n})]^{2}\neq0$ at ${\bf r}=(1,0)$, so that the positive peak has a relatively larger spectral weight. However, with increasing $u$, the two peaks move towards the Fermi energy and eventually merge into a single resonance at nearly zero energy ($\omega/\Delta_{0}$=-0.001) in the unitary limit (here $u=100t$) due to the particle-hole asymmetry of the band structure. This nearly zero-energy resonance state (ZERS) has already been observed in the Cuprate superconductors \cite{Pan}. In real heavy fermion materials, the inter-site hopping of heavy electrons, $t$, is the order of 10 K and the impurity potential is most probably greater than $u=100t\sim 0.1\,$eV, so that the scattering is likely in the unitary limit and one expects also to see a single resonance peak at nearly zero energy at the nearest-neighbor site. Fig. 2(c) plots the LDOS at the next-nearest-neighbor site. We find that the resonances are almost completely suppressed, reflecting the local nature of the impurity induced resonance state.

\begin{figure}[tbp]
\begin{center}
\includegraphics[width=1.03\linewidth]{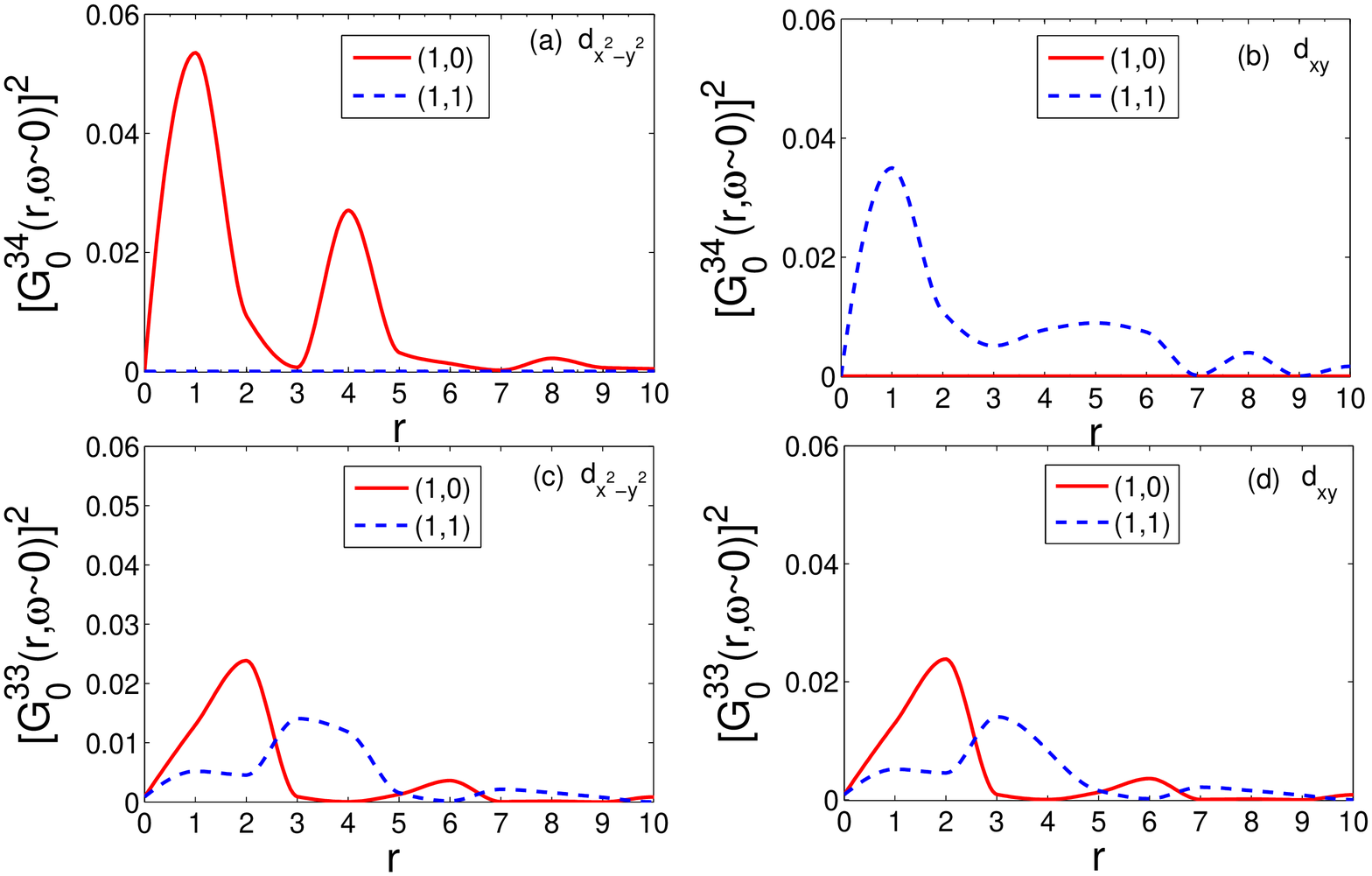}
\end{center}
\caption{(Color online) The spectral weight of $[\hat{G}^{34}_{0}({\bf r},i\omega_{n})]^{2}$ and $[\hat{G}^{33}_{0}({\bf r},i\omega_{n})]^{2}$ at the resonance energy as a function of the distance ${\bf r}$ away from the impurity for $d_{x^{2}-y^{2}}$ pairing symmetry (a) and (c), and for $d_{xy}$ (b) and (d), respectively. The red solid line denotes (1,0) direction, and blue dashed line along (1,1) direction.}
\end{figure}

\begin{figure}[tbp]
\begin{center}
\includegraphics[width=0.9\linewidth]{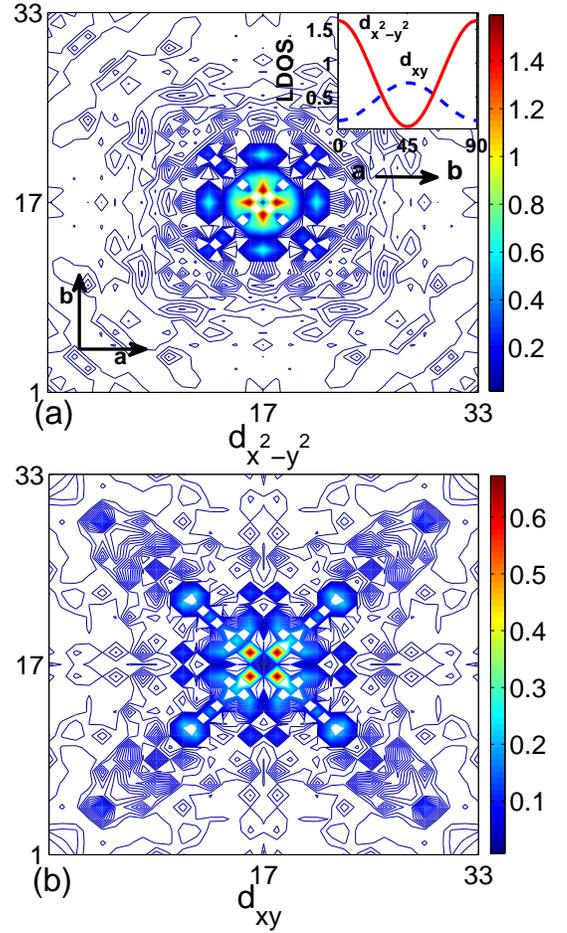}
\end{center}
\caption{(Color online) The spatial modulation of the impurity-induced resonance states in real space at the resonance energy for (a) $d_{x^{2}-y^{2}}$ and (b) $d_{xy}$ pairing states. The inset of (a) shows the height of the resonance peak for both pairing symmetry as a function of the angle from $a$-axis to $b$-axis.}
\end{figure}

Similar calculations have also been performed for $d_{xy}$ gap and the results are presented in Figs. 2(d)-2(f) for comparison. While most of the above observations are similar, there exists one striking difference: the resonances are stronger at the NNN site than the NN site for $d_{xy}$. To understand this difference, we consider the unitary limit, where only one peak exists at nearly zero energy whose spectral weight is governed mainly by $[\hat{G}^{34}_{0}({\bf r},\omega\rightarrow0)]^{2}$, so that the problem is much simplified. The $[\hat{G}^{33}_{0}({\bf r},\omega\rightarrow0)]^{2}$ term has same contributions for $d_{x^2-y^2}$ and $d_{xy}$ pairing states as shown in Figs. 3(c) and 3(d). The anomalous Green's function is related to a weighted k-space average of the gap structure, $\hat{G}^{34}_{0}({\bf r},\omega\rightarrow0)=\frac{1}{N}\sum_{\bf k}{\rm cos}(\bf k\cdot\bf{ r})\Delta^{\alpha}_{\bf k}/[(E^{\alpha}_{\bf k})^{2}+(\Delta^{\alpha}_{\bf k})^{2}]$. Because of the rapid oscillation of ${\rm cos}(\bf k\cdot\bf{ r})$ in k-space, the summation over momentum ${\bf k}$ strongly depends on the detail of the gap function. This is seen clearly in Figs. 3(a) and 3(b). For $d_{x^{2}-y^{2}}$ pairing symmetry, which has a nodal line along the $(\pm1,\pm1)$ directions, the anomalous Green's function is suppressed in this direction but has two peaks along the $(1,0)$ direction. Whereas for $d_{xy}$ gap, $\hat{G}^{34}$ is instead suppressed along the (1,0) direction but peaks at ${\bf r}=(1,1)$.

The above prediction of the nearly ZERS induced by a nonmagnetic impurity should be observed in CeCoIn$_5$ in future STM experiment. For better understanding the overall spatial pattern of the LDOS, or the local tunneling conductance probed in the STM experiment, we in Fig. 4 plots the real space spectrum on a $(33\times33)$ lattice at the resonance energy ($\omega_{res}=-0.001\Delta_{0}$) for both $d_{x^{2}-y^{2}}$ and $d_{xy}$ pairings in the unitary limit ($u=100t$). The impurity-induced resonances are localized around the impurity site. Both show a fourfold symmetry but their spatial patterns are strongly modulated by the gap structures. For $d_{x^2-y^2}$, as shown in Fig. 4(a), the maximum resonances extend  in the direction of the Ce-Ce bonds (a or b), whereas for $d_{xy}$, as shown in Fig. 4(b), the spatial pattern is rotated by $\pi/4$ and distributed mainly along the diagonal direction. The inset of Fig. 4(a) plots the height of the resonance peak as a function of the angle from $a$-axis to $b$-axis. We see a clear shift of phase in the angular dependent oscillation for the two gap functions. Interestingly, for both cases, the LDOS, or the corresponding tunneling conductance, is peaked along the antinodal direction, namely inside the superconducting gap, as required for the impurity-induced resonance, in contrast to the specific heat which is usually expected to be larger in the nodal direction where one may find more excited quasiparticles in a clean system. Given the above qualitative difference, we conclude that the spatial pattern of the tunneling conductance in the strong impurity potential scattering could provide an alternative way to determine the nodal structure of the superconducting gap function.

Finally, we have a comment on the recent STM experiment [16] where spatial structures of the in-gap states resembled the case of Ni-impurity in high-Tc cuprates, and is consistent with that induced by weak impurity scattering \cite{Zhu,Pan,Hudson,Hass,Yazdani,Salkola}. This is very different from the unitary nonmagnetic situation studied in this work. The effect of weak and magnetic impurity scattering in CeCoIn5 is currently under investigation.

In conclusion, we apply the T-matrix approach to study the effect of a single unitary nonmagnetic impurity on the electronic structures in the superconducting phase in CeCoIn$_5$. We find that impurity scattering in the unitary limit could give rise to qualitatively different spatial pattern in the LDOS measured by the STM quasiparticle interference experiment. This may provide an alternative way to determine the gap structures of the superconducting phase in CeCoIn$_{5}$ at ambient pressure. We propose that similar technique may be extended to other Ce-based heavy fermion superconductors and help to understand their gap symmetry and the underlying pairing mechanism.

This work was supported by the National Natural Science Foundation of China (NSFC) under Grants No. 11104011, Beijing Higher Education Young Elite Teacher Project under Grant No. YETP0576, Research Funds of BJTU under Grant No. 2015JBM091. Y.Y. is supported by the National 973 Project of China (Grant No. 2015CB921303), the National Natural Science Foundation of China (NSFC Grant No. 11174339) and the Strategic Priority Research Program (B) of the Chinese Academy of Sciences (Grant No. XDB07020200).

\end{document}